\documentclass{article}
\usepackage{arxiv}
\usepackage[english]{babel}
\usepackage[utf8]{inputenc} 
\usepackage[T1]{fontenc}
\usepackage{lmodern}
\usepackage{mathtools}
\usepackage{relsize}
\usepackage{amssymb}

\begin{document}
\title{Transcendental equations of the running coupling}
\author{Juuso Österman\thanks{juuso.s.osterman@helsinki.fi}\\
  University of Helsinki, Helsinki Institute of Physics}
\maketitle
\begin{abstract}
The running coupling of a generic field theory can be described through a separable differential equation involving the corresponding $\beta$-function. Only the first loop order can be solved analytically in terms of well-known functions, all further loop orders lead to transcendental equations. While obscure nowadays, many analytical methods  have been devised to study them, most specifically the Lagrange-Bürmann formula. In this article we discuss the structure of transcendental equations that take place at various loop orders. Beyond the first two loop orders, these equations are simplified by applying an optimal Pade approximant on the $\beta$-function. In general, these lead to generalizations of Lambert's equation, the solutions of which are presented in terms of a power series.  
\end{abstract}
\section{Introduction}
A renormalization group (RG) equation describes the scale dependence of a quantum field theory through the scale dependence of its parameters, e.g. mass or interaction strength (the running coupling). The relevant differential equations are constructed perturbatively in weak coupling expansion, and can in turn be integrated into transcendental equations \cite{deur,running06}. Each ascending loop order leads to more convoluted equations, with only the one-loop solution given in terms of common (yet non-algebraic) functions. 

The transcendental equations can be approached iteratively, as that leads to a unique solution with explicit scale dependence \cite{running06}. However, the iterative method only approximates the correct form of the solution \cite{nesterenko}, which motivates attempting an exact analytical evaluation process. This has been achieved up to three loops in \cite{gardi1}, by using the known solution to Lambert's equation in combination with the Padé approximation. Two further loop orders have been calculated in \cite{Kondrashuk}, where the differential equations are written in terms of an ansatz involving the Lambert W function.   

In order to consider a general loop order, we introduce a recipe utilizing (nearly) diagonal Padé approximants, which simplify the transcendental equations significantly while retaining the original accuracy. Additionally, we solve the generic transcendental equation arising from an arbitrary loop order, using the Lagrange inversion theorem. In particular, we explicitly demonstrate the procedure on the four-loop running coupling. 
\section{Background}
Coupling constants describe the strength of particular interactions in quantum field theories. The perturbative expansions in QCD and QED, in particular, are described by a single constant (respectively). The generic coupling, $x$, is defined by the renormalization process such that
\begin{equation}
x (\mu^2) Z_x(\mu^2) = x^{ur}
\end{equation}
where the right-hand side refers to the unrenormalized coupling constant, and $Z_x$ to the renormalization counterterm \cite{schwarz, running06, deur}. Through differentiation (with respect to the scale $\mu$) we find the renormalization group equation of the coupling, $x$, to be given by
\begin{equation}
\mu^2 \frac{\partial x}{\partial \mu^2} = \beta (x),
\end{equation}
where the right-hand side is called the $\beta$-function of the coupling. It can be written in terms of a perturbative power series expansion of the coupling. This explicitly reads  
\begin{equation}
\label{eq:3}
\begin{split}
\frac{\partial x}{\partial t} &= - \beta_0 x^2\sum_{n \geq 0} c_n  x^n,
\end{split}
\end{equation}
where we use the following shorthands: $t \equiv \text{ln} \mu^2$ and $c_0 = 1$. The coefficients have been evaluated up to five loops in Yang-Mills theories \cite{luthe, herzog} , QED \cite{qedbeta, Kataev}, and further in scalar theories \cite{scalarbeta}. However, the differential equations have been solved beyond these loop orders only iteratively.
\subsection{One-loop order and iterative solutions  }
As the differential equation given in equation (\ref{eq:3}) is separable, we can easily re-write it in terms of of the following integral:
\begin{equation}
\label{eq:onel}
\begin{split}
\text{ln}\frac{\mu^2}{\mu_0^2} = \int_{x_0}^{x} \frac{d \xi}{\beta( \xi)}.
\end{split}
\end{equation}
We start with the one-loop renormalization group equation (in the weak coupling expansion). Aside from the obvious purpose, this also enables us to describe the standard iterative method to evaluate further loop orders. Thus, the leading order is given by 
\begin{equation}
\begin{split}
\beta_0\text{ln}\frac{\mu^2}{\mu_0^2} &= -\int_{x_0}^{x} \frac{d \xi}{\xi^2}. \\
&= \frac{1}{x} -\frac{1}{x_0},\\
\end{split}
\end{equation}
Using trivial algebra, we can further simplify the above to
\begin{equation}
x = \frac{1}{\beta_0 \text{ln} \frac{\mu^2}{\Lambda^2}},
\end{equation}
where we have hidden all $\mu_0$-dependence such that
 \begin{equation}
\Lambda^{2 \beta_0} = \frac{\mu_0^{2\beta_0}}{e^{x_0}}.
 \end{equation}

To evaluate weak coupling at further loop levels and simultaneously avoid dealing with the non-trivial transcendental equations, one viable approach is to expand the inverse of $\beta$-function \cite{shirkov, running06, nesterenko} such that
\begin{equation}
\begin{split}
\beta_0 \text{ln}\frac{\mu^2}{\mu_0^2} &= -\int \frac{dx}{x^2(1+c_1 x+ c_2 x^2 + c_3 x^3+... )}\\
&= - \int dx \left[\frac{1}{x^2} - c_1 \frac{1}{x} -(c_2-c_1^2) - (c_3-2 c_2 c_1+c_1^3)x \right] + \mathcal{O}(x^3)\\
&= \frac{1}{x}+ c_1 \text{ln} x + (c_2-c_1^2)x + \frac{c_3-2c_1 c_2 + c_1^3}{2}x^2 + C(\mu_0)+\mathcal{O}(x^3).
\end{split}
\end{equation}
By denoting $u = \beta_0 \text{ln} \frac{\mu^2}{\Lambda^2}$ and only considering the leading order $\mathcal{O}\left(\frac{1}{x} \right)$, we obviously find the familiar one-loop solution, $x_1 = \frac{1}{u}$. However, instead of considering further cases explicitly, we write the two-loop equation such that  all terms but the inverse coupling are replaced with the one loop solution. This yields 
\begin{equation}
\begin{split}
\frac{1}{x_2} &= u -c_1 \text{ln} x_1\\
&= u+c_1 \text{ln}u.
\end{split}
\end{equation}
This process can be extended to the three-loop equation, by inserting the two-loop solution in place of the coupling such that  
\begin{equation}
\begin{split}
\frac{1}{x_3} &= u -c_1 \text{ln} x_2 -(c_2-c_1^2)x_2\\
&= u + c_1 \text{ln}(u-c_1\text{ln}u)+\frac{c_1^2-c_2}{u-c_1\text{ln} u}\\
&\simeq u + c_1 \text{ln} u + c_1^2 \frac{\text{ln} u}{u}  + \frac{c_1^2-c_2}{u} + \left(\frac{c_1^3}{2}-c_2 c_1^2 \right) \frac{\text{ln}^2 u}{u^2}+ \mathcal{O} \left( \frac{\text{ln}^3 u}{u^3}\right),
\end{split}
\end{equation}
where we performed a simple Taylor expansion on the expression at the limit $u \rightarrow \infty$. The expression at four loops can in turn be written in terms of the three-loop solution such that 
\begin{equation}
\begin{split}
\frac{1}{x_4} = u - c_1 \text{ln} x_3 + (c_1^2-c_2)x_3- \frac{c_3-2c_1 c_2 + c_1^3}{2}x_3^2,
\end{split}
\end{equation}
and the process can be continued to an arbitrary loop level. This offers a computationally efficient method with which to consider each loop order. However, it is not exact. It is our motivation to re-write the transcendental equation of each loop order such that we can analytically pursue a power series solution. To this end we introduce two methods, the Padé approximation and the Lagrange inversion theorem. 
\subsection{Padé approximation}
The approximation method is named after Henri Padé, but the idea dates back to the work of Georg Frobenius \cite{bakerpade}. In particular, the technique expresses a polynomial of order (M+N) in terms of a rational function, a Padé approximant. The corresponding approximant is denoted [N/M], using the order of the polynomials in numerator and denominator respectively:
\begin{equation}
\frac{1+\sum_{n=1}^N a_n x^n}{1+ \sum_{k= 1}^M b_k x^k} = 1+ \sum_{j=1}^{N+M} c_j x^j + \mathcal{O}(x^{N+M+1}).
\end{equation}
This approximant is unique owing to linear independence of the set $\{x^k\}$. The coefficients are determined by the following $N+M$ linear equations :
\begin{equation}
\left(1+ \sum_{j=1}^{N+M} c_j x^j \right)\left(1+ \sum_{k= 1}^M b_k x^k \right) = 1+\sum_{n=1}^N a_n x^n + \mathcal{O}(x^{N+M+1}).
\end{equation}
Order by order, these are given by
\begin{eqnarray}
c_1+ b_1 &=& a_1,\\
c_2+b_2+c_1 b_1 &=& a_2,\\
\vdots \nonumber\\
\sum_{n = 0}^{\text{min} \{M, N\}} c_{N-n} b_n &=& a_N,\\
\sum_{n=0}^{\text{min}\{N+1, M\}}c_{N+1-n}b_n &=& 0,\\
\vdots \nonumber \\
\sum_{n=0}^M c_{M+N-n} b_n &=& 0.
\end{eqnarray}
Padé approximants can be applied to simplify the integrands corresponding to each loop order. This clearly yields us more manageable transcendental equations, while the price is new higher order terms generated by the approximant. However, each loop order already obeys a similar kind of approximation, which renders the simplification viable.  The use of diagonal $[N/N]$ Padé approximants clearly simplifies the integrand most, and is additionally motivated by the reduced renormalization scale dependence. Diagonal approximants have even been shown to be scale free in the limit of large $\beta_0$ \cite{gardi2}. Also, Padé approximants have been shown to yield increasingly accurate preductions to higher order coefficients of the beta function in QCD \cite{ellis}.

This type of approximation becomes relevant at three loops and beyond, which has been known since the late 1990s. However, in terms of direct computation, it has only been extended to consider the lowest relevant order, which reduces to a widely known classic transcendental equation, Lambert's equation \cite{gardi1}. We aim to apply the Padé approximation to find the the generic structure of the transcendental equations that appear beyond three loops, in a sense a generalization of Lambert's equation. The actual evaluation takes place through the Lagrange-Bürmann formula, which enables us to write all sought-after solutions in terms of a power series.

\subsection{Lagrange Inversion theorem}
A holomorphic function can be reversed using the theorem by Joseph Lagrange (and its generalization due to Hans Bürmann), also known as the Lagrange-Bürmann theorem or the reversion of series \cite{annals8, stegun}. Let $f$ be analytic at $a$ such that $f'(a) \neq 0$, and
 \begin{equation}
 z = f(r).
 \end{equation}
 The theorem states that this equation can be solved in terms of a power series given by 
\begin{equation}
r(z) = a + \sum_{n=1}^\infty r_n \frac{[z-f(a)]^n}{n!},
\end{equation} 
where 
\begin{equation}
\label{eq:Lagrange}
r_n = \left.\frac{\partial^{n-1}}{\partial r^{n-1}} \left(\frac{r-a}{f(r)-f(a)} \right)^n\right\vert_{r = a}.
\end{equation}
The theorem can be proved using only the standard results of complex analysis \cite{whitaker}. Let us outline the steps for the expansion around origin. To any analytic $h(x)$, we can use the residual theorem to write 
\begin{equation}
\frac{1}{2 \pi i} \oint_\Psi h(x) \frac{f'(x)}{f(x)-z}dx = \sum_{k=1}^K h(x_k) 
\end{equation}
where $\{x_k\}$ is the set of solutions to the equation $f(x) = z$, which are counted with multiplicity (contained in the closed integration path $\Psi$). Supposing that $f(0) = 0 \neq f'(0)$, we can consider  $f(x)$ to be a bijection within a small circle around origin, $S$, which enables us to find a unique inverse function, $r(z) = x $ (which is also analytic). Thus, we can write
\begin{equation}
\begin{split}
r(z)&= \frac{1}{2 \pi i} \oint_S x \frac{f'(x)}{f(x)-z}dx \\
&= \frac{1}{2 \pi i} \sum_{n = 0}^\infty \oint_S x \frac{f'(x)}{f(x)} \left(\frac{z}{f(x)}\right)^n dx\\
&\equiv \sum_{n \geq 1} r_n z^n.
\end{split}
\end{equation} 
It is trivial to extract the corresponding coefficient, which simplifies to 
\begin{equation}
\begin{split}
r_n &= \frac{1}{2 \pi i}\oint_S x \frac{f'(x)}{f(x)^{n+1}}  dx\\
&= \frac{1}{n} \frac{1}{2 \pi i} \oint_S \frac{1}{f(x)^{n}}  dx\\
&= \left.\frac{\partial^{n-1}}{\partial x^{n-1}} \left(\frac{x}{f(x)} \right)\right\vert_{x = 0},
\end{split}
\end{equation}
where we applied the residual theorem. The explicit form given in equation (\ref{eq:Lagrange}) is found by re-defining the target function around $a$ instead. Specifically this means that we write for another holomorphic function $g(x)$ the following shift $g(x) = f(x)-f(a)$. This  corresponds to the same type of equation $z = g(r)$. By setting $f'(a) \neq 0$, we can again find $g(x)$ to be bijection, but unlike earlier, in a small circle around $a$. Following exactly the same steps as above, apart from defining the contour around $a$, we find the formula of interest.
\section{Exact solution through the Lambert W function}
The Lambert function dates back to 1758 when Johann Lambert solved the trinomial equation
 \begin{equation}
 x = q+x^m, 
\end{equation} 
in the sense of finding a power series corresponding to it \cite{corless, lambertorig}. This equation was in 1783 transformed to 
\begin{equation}x^{\alpha-\beta}-1 = v(\alpha-\beta) x^{\alpha},
\end{equation}
which in the limit $\beta \rightarrow \alpha =1$ becomes 
\begin{equation}
\text{ln} x = vx.
\end{equation}
Leonhard Euler calculated the power series expansion corresponding to this equation \cite{corless, eulerorig}. The solution of the symmetrized equation relates clearly to Lambert's equation,
\begin{equation}
\label{eq:lambert1}
W(z)e^{W(z)} = z,
\end{equation}
where the $W(z)$ notation is due to George Polya and Gabor Szegö \cite{szego}. By algebraic manipulation the equation by Euler becomes  
\begin{equation}
-v x e^{-v x} = -v,
\end{equation}
which in turn can be re-expressed in terms of (\ref{eq:lambert1}) such that
\begin{equation}
x = -\frac{W(-v)}{v}.
\end{equation}
The actual power series representation to the Lambert W function can be derived by using the Lagrange inversion theorem. By applying equation (\ref{eq:Lagrange}) on equation (\ref{eq:lambert1}) we find 
\begin{equation}
\begin{split}
w_n &= \left.\frac{\partial^{n-1}}{\partial w^{n-1}} \left(\frac{w}{we^{w}} \right)^n\right\vert_{w = 0}\\
&= (-n)^{n-1}.
\end{split}
\end{equation}
Thus, the power series expansion of the Lambert W function around origin is given by
\begin{equation}
W(z) = -\sum_{n=1}^\infty \frac{(-nz)^{n}}{n!n}.
\end{equation}
The generic renormalization group equation can be solved using the Lambert W function up to three loops. In the following subsections we explicitly demonstrate how the structure arises naturally from the transcendental equations of the inverse coupling \cite{gardi1}. 
\subsection{Two loops}
Let us return to the renormalization group equation of the running coupling, eq. (\ref{eq:onel}). However, this time we consider an additional term in the polynomial of the denominator. This new integral equation can be solved explicitly somewhat easily, only requiring one additional change of variables such that
\begin{equation}
\begin{split}
\beta_0\text{ln}\frac{\mu^2}{\mu_0^2} &= -\int_{x_0}^{x} \frac{d \xi}{\xi^2(1+c_1 \xi)} \\
&= \int_{x_0^{-1}}^{x^{-1}} \frac{y dy}{y + c_1}\\ 
&= \frac{1}{x} - c_1 \text{ln} \left(c_1+ \frac{1}{x} \right) + C(\mu_0),
\end{split}
\end{equation}
where we have combined all the terms arising from the lower limit of integration inside the term $C (\mu_0)$. Following the convention we used for the one-loop approximation, we choose to contain all $\mu_0$ dependence inside the logarithm on the right-hand side of the equation. Next, we write the transcendental equation in terms of the scaled inverse coupling by denoting $y = -\frac{1}{ c_1 x}$. This leads to  
\begin{equation}
\begin{split}
\text{ln} \Omega &\equiv -\frac{\beta_0}{c_1}\text{ln} \frac{\mu^2}{\Lambda^2}\\
 &= -(1-y) +\text{ln}(1-y),
\end{split}
\end{equation}
where we denote 
\begin{equation}
\frac{\beta_0}{c_1} \text{ln} \Lambda^2 = \frac{\beta_0}{c_1} \text{ln} \Lambda^2 + \frac{C(\mu_0) }{c_1} - \text{ln} c_1-1.
\end{equation}
This leads to the following transcendental equation 
\begin{equation}
\begin{split}
-\Omega &= (1-y)\text{exp}[-(1-y)]\\
&\equiv - W\left(-\Omega\right)e^{W\left(-\Omega\right)},
\end{split}
\end{equation}
where we recognize the Lambert W function. Using this expression we can re-write the inverse two-loop coupling as 
\begin{equation}
\frac{1}{x} = - c_1 \left[1+ W\left(-\Omega\right)\right].
\end{equation}
The relevant branch of Lambert W function and its behaviour has been discussed at length in literature \cite{gardi1, corless}.
\subsection{Three loops and Padé approximation}
The three-loop $\beta$ function can be described in terms of a fourth order polynomial such that 
\begin{equation}
\beta(x) = \beta_0 x^2(1+c_1 x+ c_2 x^2) + \mathcal{O}(x^5).
\end{equation}
We can achieve the same level of accuracy by writing a $[1/1]$ Padé approximant of the polynomial, following the steps given in section 2. Isolating the relevant polynomial we can write 
\begin{equation}
\begin{split}
1+ c_1 x + c_2 x^2 = \frac{1+a_1 x}{1+a_2 x} + \mathcal{O}(x^3),
\end{split}
\end{equation}
where the new coefficients correspond to the following set of linear equations
\begin{eqnarray}
c_1 + a_2 &=& a_1,\\
c_2 + c_1 a_2 &=& 0, 
\end{eqnarray}
which in turn yield
\begin{eqnarray}
a_1 &=& c_1-\frac{c_2}{c_1},\\
a_2 &=& -\frac{c_2}{c_1}.
\end{eqnarray}
Through this procedure we have reduced the integrand to the same general structure as the one we found at the two-loop order. In particular we avoid a more convoluted variant of the transcendental equation with a higher order polynomials. The relevant structures are discussed at more length in section 4.

By inserting the Padé approximant to the integral representation of the three-loop equation, we find
\begin{equation}
\begin{split}
\beta_0\text{ln}\frac{\mu^2}{\mu_0^2} &= -\int_{x_0}^{x} \frac{(1+a_2 \xi) d \xi }{\xi^2(1+a_1 \xi)} \\
&= \int_{x_0^{-1}}^{x^{-1}} \frac{(y+a_2) dy}{y + a_1}\\ 
&= \frac{1}{x} +(a_2-a_1) \text{ln} \left(a_1+ \frac{1}{x} \right) + C(\mu_0).
\end{split}
\end{equation}
We continue by moving the integration constant inside the logarithm and writing the transcendental equation in terms of the scaled inverse coupling $y = -\frac{1}{(a_2-a_1)x}$. Here we have assumed that the coefficients have non-trivial values, as $a_1 = a_2$ would lead to the one-loop solution. Thus, the transcendental equation is given by
\begin{equation}
\begin{split}
\text{ln} \Omega &\equiv \frac{\beta_0}{a_2-a_1}\text{ln}\frac{\mu^2}{\Lambda^2}\\
  &= \left(\frac{a_1}{a_2-a_1} - y \right) + \text{ln} \left( \frac{a_1}{a_2-a_1}-y\right).\\
\end{split}
\end{equation}
Similar to the two-loop transcendental equation, we can easily discern the logarithmic version of Lambert's equation. Thus, we are able to write the inverse coupling in terms of the Lambert W function such that
\begin{equation}
\begin{split}
\frac{1}{x} &= - \{a_1 - (a_2-a_1)W(\Omega)\}\\
&=  -\left\{c_1-\frac{c_2}{c_1}+c_1 W(\Omega) \right\}.
\end{split}
\end{equation}

\section{General structure of transcendental equations}
Already at the two-loop order we witnessed the appearance of a highly non-trivial transcendental equation. At the three-loop level, we applied the Padé approximation to retain the same level of complexity. However, upon considering further loop orders, or even the three-loop order without algebraic manipulation, it is obvious that we will encounter even more involved transcendental equations, which do not reduce to Lambert's equation.  We choose to demand a small amount of regularity from the rational function integrand, in particular binding the coefficients of the $\beta$-function in a specific manner, if need be considering a different number of active fermions. In order to demonstrate the additional condition we write the (N+1)-loop integral equation in terms of the inverse coupling $y$ such that
\begin{equation}
\label{eq:pureb}
\begin{split}
\text{ln} \Omega &= \int dx  \frac{1}{\beta_N(x)}\\
&= \int dy \frac{y^N}{y^N + \sum_{k=1}^{N} b_k y^{N-k}}\\
&= \int dy \frac{y^N}{\prod_{k=1}^N (y+a_k)}.
\end{split}
\end{equation}
Here we write the denominator as a product of $n$ monomials, which in turn implies that no two roots are allowed to be equal. This condition is applied to all integrals in this section, to pedagogically motivate the use of diagonal or nearly diagonal Padé approximants. However, as far as results are concerned, the coefficients of the $\beta$-function are limited only such that these Padé approximants contain a denominator that decomposes in the manner described above.

The integral given in equation (\ref{eq:pureb}) is rather straightforward to solve; the relevant steps of the derivation are given in the appendix A. Next we absorb the lower limit on the right-hand side to $\text{ln}\tilde{\Omega} = \text{ln} \Omega - C$. Then we  write the logarithmic representation of the corresponding transcendental equation such that
\begin{equation}
\text{ln} \tilde{\Omega} =  y +\sum_{n=1}^N \tilde{c}_n \text{ln}(y+a_n),
\end{equation}
the preferable form of which is given by
\begin{equation}
\label{eq:proood}
\tilde{\Omega}e^{-y} = \prod_{n=1}^N  (y+a_n)^{\tilde{c}_n} .
\end{equation}
This result corresponds to the [N/0] Padé approximant of the denominator, i.e. the original polynomial. The structural opposite of this being the [0/N] Padé approximant, which leads to the following integral equation
\begin{equation}
\begin{split}
\text{ln} \Omega &= -\int \frac{dx}{x^2} \sum_{k=0}^N c_k x^k\\
&= \int dy \sum_{k=0}^N \frac{c_k}{y^k}.
\end{split}
\end{equation}
Following similar steps to those given above, we find the corresponding transcendental equation to be
\begin{equation}
\tilde{\Omega}\text{exp}\left( -y + \sum_{k=1}^{N-1} \frac{c_{k+1}}{k y^k} \right) = y^{c_1},
\end{equation}
which looks very uninviting to solve for any $N \geq 2$. The former is obviously the  preferable form to consider analytically, as the inversion theorem operates through derivatives. However, we are able to reduce the transcendental equation further. This is achieved through diagonal, or nearly diagonal Padé approximants. Supposing that $N = 2n$, we can obviously write instead the $[n/n]$ Padé approximant to achieve 
\begin{equation}
\text{ln} \Omega = \int dy \frac{y^n + \sum_{k=0}^{n-1} c_k y^k}{y^n + \sum_{j=0}^{n-1} b_j y^j}.
\end{equation}
Using again the appendix A, we can immediately write the result in the following form
\begin{equation}
\label{eq:easy1}
\tilde{\Omega}e^{-y} =  \prod_{n=1}^n  (y+a_n)^{\tilde{c}_n},
\end{equation}
which halves the generalized polynomial structure on the right-hand side of equation (\ref{eq:proood}).

In case the loop order is even (the polynomial is of odd order), we can instead write $m = \lfloor N/2 \rfloor$ and instead consider the $[m/m+1]$ Padé approximant to find 
\begin{equation}
\begin{split}
\text{ln} \Omega = \int dy \left(\frac{y^m + \sum_{k=0}^{m-1} c_k y^k}{y^m + \sum_{j=0}^{m-1} b_j y^j}+\frac{c_m}{y\left(y^m + \sum_{j=0}^{m-1} b_j y^j \right)} \right),
\end{split}
\end{equation} 
which in turn yields a variant of the diagonal result of the order $m+1$ such that
\begin{equation}
\label{eq:easy2}
\tilde{\Omega}e^{-y} = y^{\hat{c}_0}\prod_{k=1}^m (y+a_k)^{\tilde{c}_k}.
\end{equation} 
The results given in equations (\ref{eq:easy1}) and (\ref{eq:easy2}) generalize Lambert's equation, and can be solved through the Lagrange inversion theorem (with added combinatorial difficulty). 

\section{Four loops and  Padé approximation}
After having established the types of transcendental equations we are interested in, it is logical to apply the tools we have to the higher order cases, which have so far been left ambiguous. In particular, we start with the four-loop approximation of the $\beta$-function.  As indicated in the earlier section, we aim to avoid higher order contributions by initiating our approach with the $[1/2]$ Padé approximant of the following polynomial
\begin{equation}
1 + c_1 x + c_2 x^2 + c_3 x^3 = \frac{1+ a_1 x}{1+ a_2 x + a_3 x^2} + \mathcal{O}(x^4),
\end{equation}
where we find the following linear relations for the parameters 
\begin{eqnarray}
a_2+c_1 &=& a_1,\\
c_1 a_2 + c_2 + a_3 &=& 0,\\
c_1 a_3 + c_3 &=& 0,
\end{eqnarray}
which define the Padé approximant uniquely such that
\begin{eqnarray}
a_1 &=& c_1- \frac{c_2}{c_1}+\frac{c_3}{c_1^2},\\
a_2  &=& -\frac{c_2}{c_1}+\frac{c_3}{c_1^2},\\
a_3  &=& -\frac{c_3}{c_1}.
\end{eqnarray}
These can be used to re-write the four-loop approximation of the renormalization group equation as
\begin{equation}
\begin{split}
\beta_0\text{ln}\frac{\mu^2}{\mu_0^2} &= -\int_{x_0}^{x} \frac{(1+a_2 \xi + a_3 \xi^2) d \xi }{\xi^2(1+a_1 \xi)} \\
&= \int_{x_0^{-1}}^{x^{-1}}dy \left(1 +\frac{a_2-a_1 - \frac{a_3}{a_1}}{y+ a_1} + \frac{a_3}{a_1 y}  \right)\\ 
&= y +\left(a_2-a_1-\frac{a_3}{a_1} \right) \text{ln} \left(a_1+ y \right) + \frac{a_3}{a_1}\text{ln} y + D(\mu_0),
\end{split}
\end{equation}
where we denote $y \equiv \frac{1}{x}$. Let us simplify the notation further by writing
\begin{eqnarray}
A &=&  a_2 - a_1 -\frac{a_3}{a_1},\\
B &=& \frac{a_3}{A a_1},\\
C &=& \frac{a_1}{A},\\
z &=& \frac{y}{A}.
\end{eqnarray}
Thus, we are able to write the logarithmic representation of the transcendental equation as
\begin{equation}
\begin{split}
\label{eq:abbreviated}
\text{ln} \Omega &\equiv \frac{\beta_0}{A}\text{ln}\frac{\mu^2}{\Lambda^2}\\
&= z + \text{ln} \left(C + z\right)+ B \text{ln} z. 
\end{split}
\end{equation}
This result indeed resembles the type of transcendental equation taking place upon applying an almost diagonal Padé approximation, given in equation (\ref{eq:easy2}):
\begin{equation}
\label{eq:abbreviated2}
\begin{split}
(C+z)z^B = \Omega e^{-z}.
\end{split}
\end{equation} 
\subsection{Inversion at four loops}
 The equation of interest can be rephrased conveniently such that
\begin{equation}
(v-a)v^b = c e^{-v}.
\end{equation}
We approach this new problem by denoting $f(v) = \frac{e^{-v}}{v^b}$, using which the equation transforms to   
\begin{equation}
\label{eq:standard}
v = a +c f(v).
\end{equation}
By re-organizing these terms, we re-write the equation in terms of a constant valued function
\begin{equation}
F(v) = v- c f(v) \equiv a.
\end{equation}  
We choose to expand around the bound value $a$, and obviously $F'(a) = -c f'(a) \neq 0$, which explicitly justifies the use of the Lagrange inversion theorem. In order to simplify the power series solution, we introduce three helpful algebraic relations:
\begin{eqnarray}
F(a)-a &=& -c f(a),\\
v-a &=& c f(v),\\
F(v)-F(a) &=& F(v)-a + cf(a)= c f(a).
\end{eqnarray}
The sought after power series solution via the inversion formula is given by
\begin{equation}
v(a) = a + \sum_{k \geq 1} \frac{v_k}{k!} [a-F(a)]^k,
\end{equation}
where we again define the power series coefficient as 
\begin{equation}
\begin{split}
v_k &= \left.\partial_v^{k-1} \left( \frac{v-a}{F(v)-F(a)}\right)^k \right|_{v= a}\\
&= \left. \left[\frac{1}{ f(a)}\right]^k \partial_v^{k-1} f(v)^k \right|_{v= a}.
\end{split}
\end{equation}
Thus, we find the following simpler form of the sought after solution to the transcendental equation
\begin{equation}
\label{eq:simple}
v = a + \sum_{k \geq 1} \left.  \frac{c^k}{k!}  \partial_v^{k-1} f(v)^k \right|_{v= a},
\end{equation}
which is in agreement with the result presented in \cite{Mugnaini}.
Let us next consider the derivative more explicitly by writing 
\begin{equation}
\begin{split}
\partial_v^{k-1} f(v)^k &= \partial_v^{k-1} \frac{e^{-kv}}{v^{kb}}\\
&= \sum_{j = 0}^{k-1} {k-1 \choose j} \left[ \frac{\partial^j}{\partial v^j} e^{-k v} \right] \left[ \frac{\partial^{k-1-j}}{\partial v^{k-1-j}} v^{-kb}\right]\\
&= (-1)^{k-1} \sum_{j = 0}^{k-1}  {k-1 \choose j} \frac{k^j}{kb-1} \prod_{l=0}^{k-1-j} (kb-1+l) \frac{e^{-kv}}{v^{kb+k-1-j}}.
\end{split}
\end{equation}
Thus, we find the following solution to our transcendental equation
\begin{equation}
\begin{split}
v = a - \sum_{k \geq 1}\sum_{j = 0}^{k-1}  \frac{(-c)^k k^j}{(k-1-j)! j! k (kb-1)  }  \prod_{l=0}^{n-1-j} (kb-1+l) \frac{e^{-ka}}{a^{kb+k-1-j}}.
\end{split}
\end{equation}
By introducing the Pochhammer symbol such that 
\begin{equation}
(v)_k = \frac{\Gamma(v+k)}{\Gamma (k)},
\end{equation}
we can simplify the solution given above to the more aesthetic
\begin{equation}
\begin{split}
v = a - \sum_{k \geq 1}\sum_{j = 0}^{k-1}  \frac{(-c)^k k^j}{(k-1-j)! j! k  }   (kb)_{k-1-j} \frac{e^{-ka}}{a^{kb+k-1-j}}.
\end{split}
\end{equation}
\subsection{Full expression at four loops}
Let us return to the relation given in equation (\ref{eq:abbreviated2}). Applying this to the previous identity, we find
\begin{equation}
\begin{split}
z = -C - \sum_{k \geq 1}\sum_{j = 0}^{k-1}  \frac{(-\Omega)^k k^j}{(k-1-j)! j! k  }   (kB)_{k-1-j} \frac{e^{kC}}{(-C)^{kB+k-1-j}},
\end{split}
\end{equation}
or in terms of the inverse coupling
\begin{equation}
\label{eq:fullfour}
y = - AC -A\sum_{k \geq 1}\sum_{j = 0}^{k-1}  \frac{\left[-\left(\frac{\mu^2}{\Lambda^2} \right)^\frac{\beta_0}{A}\frac{e^{C}}{(-C)^{B+1}}\right]^k }{(k-1-j)! j! k^2 }  \left(kB \right)_{k-1-j} (-kC)^{j+1}.
\end{equation}
It is noteworthy that (\ref{eq:abbreviated}) is not a unique representation, and therefore, we could have instead written 
\begin{eqnarray}
A &=& \frac{a_3}{a_1},\\
B &=& \frac{ a_2 - a_1 -\frac{a_3}{a_1}}{A},\\
C &=& \frac{a_1}{A},\\
\tilde{z} &=& \frac{y}{A},
\end{eqnarray}
to introduce instead
\begin{equation}
\frac{\beta_0}{A}\text{ln}\frac{\mu^2}{\Lambda^2} = \tilde{z} + B \text{ln} \left(C + \tilde{z}\right)+  \text{ln} \tilde{z}. 
\end{equation}
This, in turn, would lead us to find a solution around $z=0$ and to consider the function $f(x) = \frac{e^{-x}}{(x+a)^b}$. The resulting power series solution is given by
\begin{equation}
y = -A\sum_{k \geq 1}\sum_{j = 0}^{k-1}  \frac{\left[-\left(\frac{\mu^2}{\Lambda^2} \right)^\frac{\beta_0}{A}\frac{1}{(-C)^{B+1}}\right]^k }{(k-1-j)! j! k^2 }  \left(kB \right)_{k-1-j} (-kC)^{j+1}.
\end{equation}
Also, we note that the four-loop approximation did not demand any new imposed  conditions on the denominator structure of the integrand. Rather, the only notable difference to the lower loop order cases was the well defined additional logarithmic term. However, the arising power series solution is somewhat more involved. Also, the former result, given in equation (\ref{eq:fullfour}), is more attractive as the proper parameter values would make it exponentially suppressed (in particular $C < 0$).
\section{Discussion on the higher loop orders}
As elaborated in section 4, the preference for diagonal Padé approximants leads us to consider the higher order integrands in terms of rational functions, the denominators of which decompose to a product of n separate monomials. Specifically, in order to optimize our efforts, we limit to Padé approximants of types $[n-1/n]$ or $[n/n]$, which lower the degree of the sought after polynomial and thus greatly simplify the transcendental equation of interest. In this manner we can find the five-loop coupling structure given by the $[2/2]$  Padé approximant such that
\begin{equation}
\begin{split}
\text{ln} \Omega &= \int dy \frac{1+\frac{a_1}{y}+ \frac{a_2}{y^2}}{1+ \frac{a_3}{y}+\frac{a_4}{y^2}}\\
&= \int dy  \left[1 + \frac{(a_1-a_3)y}{y^2+a_3 y+ a_4}+ \frac{a_2-a_4}{y^2+a_3 y + a_4 }  \right].
\end{split}
\end{equation}
By recognizing the decomposition of the denominator in terms of the roots of any second order polynomial, we find  $y^2+a_3 y + a_4 = (y+a_{+})(y+a_{-})$  and $a_{\pm} = \frac{a_3\pm \sqrt{a_3^2-4a_4}}{2}$. This leads to the following logarithmic representation of the transcendental equation

\begin{equation}
\text{ln} \tilde{\Omega} = y + \frac{a_1-a_3}{a_+ - a_-} \left[\text{ln}\frac{(y+a_+)^{a_+}}{(y+a_-)^{a_-}}\right] +\frac{a_2-a_4}{a_+ - a_-} \text{ln} \left( \frac{y+a_-}{y+a_+} \right)
\end{equation}  
These lead to the following type of transcendental equation under sufficient conditions
\begin{equation}
\Xi e^{-y} = (y+a_+)^A (y+a_-)^B,
\end{equation}
or given in a similar manner as in the previous section 
\begin{equation}
v = a + b \frac{e^{-v}}{(v+c)^d},
\end{equation}
which leads us to the realization that the five-loop transcendental equation generalizes the four-loop case, something that we recognize taking place also with three and two loops. This corresponds to the pattern we described in section 4, with each consecutive even and odd loop orders, the approximants $[n-1/n]$ and $[n/n]$, being effectively the same (apart from parameter values). Thus, let us establish the duality between the transcendental equations and the n-loop order approximations of the coupling. This allows us to clearly state that all structural information is demonstrated by the odd loop orders, i.e. the diagonal approximants. 

Let us consider an arbitrary $(N+1)$-loop (odd) order transcendental equation and simplify it, denoting $n = \frac{N}{2}$, to the familiar structure such that 
\begin{equation}
v = a + b \frac{e^{-v}}{\Pi_{k= 1}^{n-1} (v+a_k)^{c_k}}.
\end{equation}
This type of equation can be solved following the exact same steps as before. However, this time we encounter a multinomial coefficient upon derivating instead of the binomial. Explicitly we can formulate the transcendental equation of interest such that the function $f$ from the previous section becomes
\begin{equation}
f(v) = \frac{e^{-v}}{\Pi_{k= 1}^{n-1} (v+a_k)^{c_k}},
\end{equation}
and therefore we find the following derivative in the power series coefficient
\begin{equation}
\begin{split}
\partial_v^{k-1} f(v)  &= \sum_{j_1+j_2+...+j_n =k-1} {k-1 \choose j_1, j_2,....j_n} \left[ \frac{\partial^{j_n}}{\partial v^{j_n}} e^{-kv}\right]\prod_{l=1}^{n-1} \left[ \frac{\partial^{j_l}}{\partial v^{j_l}} (v+a_l)^{-kc_l} \right] \\
&= (-1)^{k-1} \sum_{j_1+j_2+...+j_n =k-1} {k-1 \choose j_1, j_2,....j_n} \left[ k^{j_n}  e^{-kv}\right]\prod_{l=1}^{n-1} \left(kc_l \right)_{j_l} (v+a_l)^{-kc_l-j_l}. 
\end{split}
\end{equation}
Thus, applying above results to equation (\ref{eq:simple}), we find the generic solution type to the arbitrary (diagonal) loop order inverse coupling
\begin{equation}
\label{eq:final}
v = a - \sum_{k=1}^\infty \sum_{j_1+j_2+...+j_n =k-1} \frac{(-1)^k}{k!} {k-1 \choose j_1, j_2,....j_n} \left[ k^{j_n}  e^{-ka}\right]\prod_{l=1}^{n-1} \left(kc_l \right)_{j_l} (a+a_l)^{-kc_l-j_l}.
\end{equation}
The resulting series converges very fast with sufficiently large values of the parameters, and thus offers a viable alternative to the iterative approach. Of course, it is noteworthy that some of the uniqueness is lost by approaching the problem through transcendental equations in earnest, as the diagonal structure we described can be chosen to be expanded around multiple separate root values. However, this ambiguity is problematic in appearance only, as a suitable physical value, $a \gg 1$, can be chosen (if any such exists) from the set of roots, leading to correct physical behaviour at the weak coupling limit.
\section{Conclusions}
In this work, we have evaluated the inverse four-loop running coupling analytically for a generic quantum field theory. The result is given in terms of a power series, which solves a generalized variant of Lambert's equation. The components of the power series solution, given in equation (\ref{eq:fullfour}), are characterized by an exponential dependence on the parameters of the transcendental equation. This implies fast convergence for a suitable set of physical parameter values.   

We have also discussed the general structure of the n-loop transcendental equations. The equations of interest are generated by the subset of (nearly) diagonal Padé approximants, the denominators of which decompose into a product of discrete roots. We have noted that these transcendental equations are generalizations of Lambert's equation, with each ascending loop order adding new generalized polynomials into the structure. We also established that the forms of the transcendental equations are fully described by the odd loop orders, given in equation (\ref{eq:easy1}). 

This generic transcendental equation of an arbitrary loop order has been solved analytically using the Lagrange inversion theorem. The power series solution is akin to the four-loop solution, in terms of its exponential behaviour, as seen in equation (\ref{eq:final}). However, it adds a combinatorially increasing number of components corresponding to the loop order. 

\section*{Acknowledgements}
The author wishes to thank Antti Kupiainen, Anca Tureanu and Aleksi Vuorinen for enlightening discussions and feedback. This work has been supported by the European Research Council, grant no.~725369, and by the Academy of Finland, grant no.~1322507. 
\appendix
\section{Reduction identities}
In this appendix we give a more detailed description about the integral identities used in sections 4-6. The rational integrals are structured such that the denominators decompose into a product of $n$ monomials, whereas the numerators are taken to be polynomials of arbitrary orders. The integrals are carried out in terms of the inverse coupling, $y$, following the notation in section 4. Also, throughout this appendix we neglect to write the integration constants explicitly.

Let us start with the case of the numerator being a monomial of lower order. In the extreme case we note that
\begin{equation}
\begin{split}
\int dy \frac{y^{n-1}}{\prod_{k=1}^n(y+a_k)} &= \int dy \left[\frac{1}{y+a_1}- \frac{p_{n-2}(y)}{\prod_{k=1}^n(y+a_k)} \right]\\
&= \text{ln} (y+a_1) - \sum_{j=1}^{n-2} a_j \int dy \frac{y^j}{\prod_{k=1}^n(y+a_k)},
\end{split}
\end{equation} 
where we denote $p_k(y)$ as a $k$th order polynomial with the following condition 
\begin{equation}
y^{n-1}+ p_{n-2}(y) = \prod_{k=2}^n(y+a_k).
\end{equation}
Next, we consider a monomial of the order $0 <m < n $. Following similar steps to those given above, we find
\begin{equation}
\label{eq:ym}
\begin{split}
\int dy \frac{y^{m}}{\prod_{k=1}^n(y+a_k)} &= \int dy \left[\frac{1}{\prod_{k=1}^{n-m}(y+a_k)}- \frac{p_{m-1}(y)}{\prod_{k=1}^n(y+a_k)} \right].\\
\end{split}
\end{equation}
In order to solve the first part of equation (\ref{eq:ym}), we wish to decompose it such that each root is separately summed. This leads to $n-m$ linear equations which uniquely determine the values of the coefficients (up to scaling) such that
\begin{eqnarray}
\sum_{k = 0}^{n-m} A_k &=& 0,\\
\sum_{j=0}^{n-m} \sum_{k=0}^{n-m} A_j (1 -\delta_{jk}) a_k &=& 0,\\
\vdots \nonumber \\
\sum_{j=0}^{n-m} \sum_{l \neq j}^{n-m}  A_j \prod_{k \neq j, l} a_k &=& 0.
\end{eqnarray}
This provides us with the following decomposition 
\begin{equation}
\label{eq:decomp}
\begin{split}
\int dy \frac{1}{\prod_{k=1}^{n-m}(y+a_k)} &= \frac{1}{\sum_{j=0}^{n-m}   A_j \prod_{k \neq j} a_k}\sum_{l=1}^{n-m} \int dy \frac{A_l}{y+a_l}\\
&=\sum_{l=1}^{n-m} \frac{A_l \text{ln} (y+a_l)}{\sum_{j=0}^{n-m}   A_j \prod_{k \neq j} a_k}.
\end{split}
\end{equation}
Thus, we can write for any $y^m$
\begin{equation}
\begin{split}
\int dy \frac{y^{m}}{\prod_{k=1}^n(y+a_k)} &= \sum_{j=1}^{n-m} c_j \text{ln}(y+a_j) - \int dy \frac{p_{m-1}(y)}{\prod_{k=1}^n(y+a_k)} \\
&= \sum_{j=1}^n \tilde{c}_j \text{ln}(y+a_j), 
\end{split}
\end{equation}
where we iterated the earlier result $m$ times and hid all contributions from the lower order polynomials within the set of new constants $\{\tilde{c}_j \}$. Using this result, we can consider any polynomial $p_n(y)= \sum_{k=0}^n c_{k}y^{n-k}$ in the numerator:
\begin{equation}
\begin{split}
\int dy \frac{p_n (y)}{\prod_{k=1}^n(y+a_k)} &=c_0 \int dy + \int dy \frac{\tilde{p}_{n-1}}{\prod_{k=1}^n(y+a_k)}\\
&= c_0 y + \sum_{k=1}^n \tilde{c}_k \text{ln} (y+a_k).
\end{split}
\end{equation}
This result describes a slight generalization of the integral we need to consider with the diagonal $[n/n]$ Padé approximants.

Additionally, we want to consider the integrals related to the non-diagonal Padé approximants. Thus, we introduce an integrand of the original coupling, in which the numerator is a higher order polynomial than the denominator. It is trivial to re-write this kind of integral in terms of the inverse coupling and to isolate the unfamiliar parts:
\begin{equation}
\begin{split}
\int dy \frac{p_{n+m}(y)}{y^m \prod_{k=1}^n(y+a_k)} &= \sum_{j = 1}^{m}  \int dy \frac{d_j}{y^j\prod_{k=1}^n(y+a_k)} + \int dy \frac{\tilde{p}_n (y)}{\prod_{k=1}^n(y+a_k)}. \\
\end{split}
\end{equation}
The unknown integrals, characterised by the denominator of the order $n+j$, can be decomposed in similar manner to equation (\ref{eq:decomp}). By applying some linear algebra, we find
\begin{equation}
\begin{split}
\int dy \frac{1}{y^j\prod_{k=1}^n(y+a_k)} &= \sum_{k = 1}^n \int dy \frac{c_k}{y^j(y+a_k)}\\
&= \sum_{k = 1}^n \int dy \frac{c_k}{a_k y^{j-1}}\left[\frac{1}{y}-\frac{1}{y+a_k} \right]\\
&= \sum_{k = 1}^n \frac{c_k}{a_k} \int dy \left[\sum_{l=0}^{j-1}\frac{1}{(-a_k)^{l} y^{j-l}} -(-a_k)^{1-j}\frac{1}{y+a_k}\right]\\
&= \sum_{k = 1}^n c_k\left[\sum_{l=1}^{j-1} \frac{(-a_k)^{-l}}{l y^{l}} -(-a_k)^{-j} \text{ln} y + (-a_k)^{-j} \text{ln} (y+a_k) \right],
\end{split}
\end{equation}
where we use the convention that the summation vanishes completely if the maximum index is smaller than the minimum. This allows us to identify the structure of the full integral as 
\begin{equation}
\int dy \frac{p_{n+m}(y)}{y^m \prod_{k=1}^n(y+a_k)} = \sum_{j=1}^{m-1} \frac{\tilde{c}_j}{y^j} + \tilde{c}_0 \text{ln} y+\tilde{e}_0 y + \sum_{k = 1}^n \tilde{v}_k \text{ln} (y + a_k).
\end{equation}
In particular we find the $[n/n+1]$ Padé approximants to be described by the following special case
\begin{equation}
\int dy \frac{p_{n+1}(y)}{y\prod_{k=1}^n(y+a_k)} = \tilde{c}_0 \text{ln} y+\tilde{e}_0 y + \sum_{k = 1}^n \tilde{v}_k \text{ln} (y + a_k).
\end{equation}
And as a last case of minor interest, we list the extreme case of $[0/n]$ Padé approximant, which yields the following type of integral
\begin{equation}
\begin{split}
\int dy \left(1 + \sum_{k=1}^n a_k y^{-k}  \right) =  y - \sum_{k=1}^{n-1} \frac{a_{k+1}}{k y^k} + a_1\text{ln} y.  
\end{split}
\end{equation}
\bibliographystyle{unsrt}
\bibliography{referencespade}
\end{document}